\documentclass[twocolumn,journal]{IEEEtran}
\usepackage[latin9]{inputenc}
\usepackage{pifont}
\usepackage{amsmath}
\usepackage{amssymb}
\usepackage{graphicx}

\makeatletter
\IEEEoverridecommandlockouts

\usepackage{color}
\usepackage{bm}
\usepackage{algorithm}
\usepackage{algorithmic}
\usepackage{multirow}
\usepackage{booktabs}
\usepackage{array}
\usepackage{amsthm}
\usepackage{lipsum}
\usepackage{enumerate}
\usepackage{stfloats}
\usepackage{subfigure}
\usepackage{cases}
\usepackage{diagbox}
\usepackage{threeparttable}
\usepackage{cite}
\usepackage{pifont}
\usepackage{xcolor}
\usepackage{xpatch}

\makeatother

\begin{document}
\title{Reconfigurable Intelligent Surfaces 2.0: Beyond Diagonal Phase Shift
Matrices}
\author{Hongyu Li, \IEEEmembership{Student Member, IEEE,} Shanpu Shen, \IEEEmembership{Senior Member, IEEE,}\\
Matteo Nerini, \IEEEmembership{Student Member, IEEE,} and Bruno Clerckx, \IEEEmembership{Fellow, IEEE}
\thanks{\textit{(Corresponding author: Shanpu Shen.)}}
\thanks{H. Li and M. Nerini are with the Department of Electrical and Electronic
Engineering, Imperial College London, London SW7 2AZ, U.K. (e-mail: \{c.li21,m.nerini20\}@imperial.ac.uk).}
\thanks{S. Shen is with the Department of Electrical Engineering and Electronics, University of Liverpool, Liverpool, L69 3GJ, U.K. (e-mail: Shanpu.Shen@liverpool.ac.uk).}
\thanks{B. Clerckx is with the Department of Electrical and Electronic Engineering, Imperial College London, London SW7 2AZ, U.K. and with Silicon Austria Labs (SAL), Graz A-8010, Austria (e-mail: b.clerckx@imperial.ac.uk).}}

\maketitle
\begin{abstract}
Reconfigurable intelligent surface (RIS) has been envisioned as a promising technique to enable and enhance future wireless communications due to its potential to engineer the wireless channels in a cost-effective manner. 
Extensive research attention has been drawn to the use of conventional RIS 1.0 with diagonal phase shift matrices, where each RIS element is connected to its own load to ground but not connected
to other elements. However, the simple architecture of RIS 1.0 limits its flexibility of manipulating passive beamforming. 
To fully exploit the benefits of RIS, in this paper, we introduce RIS 2.0 beyond diagonal phase shift matrices, namely beyond diagonal RIS (BD-RIS). 
We first explain the modeling of BD-RIS based on the scattering parameter network analysis and classify BD-RIS by the mathematical characteristics of the scattering matrix, supported modes, and architectures. 
Then, we provide simulations to evaluate the sum-rate performance with different modes/architectures of BD-RIS. 
We summarize the benefits of BD-RIS in providing high flexibility in wave manipulation, enlarging coverage, facilitating the deployment, and requiring fewer resolution bits and scattering elements.
Inspired by the benefits of BD-RIS, we also discuss potential applications of BD-RIS in various wireless systems. 
Finally, we list key challenges in modeling, designing, and implementing BD-RIS in practice and point
to possible future research directions for BD-RIS. 
\end{abstract}


\vspace{-0.5 cm}

\section{Introduction}

Wireless networks for the first five generations have been operated by catering the uncontrollable wireless environment through various sophisticated designs at the transmitter/receiver. For beyond 5G and 6G, however, wireless networks are expected to have manipulations of both transmitter/receiver and wireless environment, thanks to the emergence of a promising technique, namely reconfigurable intelligent surface (RIS) \cite{di2020smart,wu2019towards}. 
RIS consists of numerous passive reconfigurable scattering elements so that it can manipulate the wireless environment and thus enhance the spectrum and energy efficiency of the wireless network.
The advantages of RIS have been demonstrated in various wireless systems, such as enabling integrated sensing and communication and improving power relaying \cite{wu2021intelligent}.
However, most existing works focus on using a simple RIS model with diagonal phase shift matrix, here referred to as RIS 1.0, where each RIS element is connected to its own reconfigurable impedance without inter-element connections. More specifically, there are two limitations of conventional RIS 1.0: 1) Conventional lossless RIS 1.0 can only control the phase of incident signal, which limits capability for manipulating passive beamforming and thus degrades the performance. 2) It only enables the signal reflection towards the same side, which limits the coverage.

To address these limitations of RIS 1.0 and further enhance the performance gain of RIS, in this paper, we branch out to RIS 2.0 by introducing inter-element connections at the expense of additional circuit complexity, whose mathematical model is not limited to be diagonal matrices. We refer to this RIS 2.0 as beyond diagonal RIS (BD-RIS). 
We start from the BD-RIS modeling through scattering parameter network analysis. Then, we classify the BD-RIS based on the characteristics of the BD-RIS matrix, the supported modes, and the architectures, and categorize the existing BD-RIS mode/architecture design works \cite{shen2021,li2022dynamic,li2022reconfigurable,li2022,li2022beyond} accordingly.
Next, we evaluate the achievable sum-rate performance for a multi-user system with different modes/architectures of BD-RIS using the beamforming design algorithms proposed in \cite{li2022,li2022beyond}. We summarize the benefits of BD-RIS such as high flexibility in wave manipulation and full-space coverage and review corresponding BD-RIS works \cite{santamaria2023snr,wang2023channel,bartoli2023spatial,nerini2021reconfigurable}.
Inspired by the benefits of BD-RIS, we look ahead to potential applications of BD-RIS in future wireless networks, such as enabling energy-efficient power relay in the power grid, and assisting wireless sensing in vehicular networks. We also discuss key challenges and future work of BD-RIS. Finally, we conclude this paper.

\vspace{-0.3 cm}

\section{Modeling and Classification of BD-RIS}

In this section, we introduce the model of BD-RIS based on the scattering parameter network analysis, and classify BD-RIS based on different modes and architectures.

\vspace{-0.3 cm}

\subsection{BD-RIS Model}

An $M$-element RIS is a passive device modeled as $M$ antennas connected to an $M$-port reconfigurable impedance network \cite{shen2021}.
The $M$-port reconfigurable impedance network is constructed by reconfigurable passive components and mathematically characterized by the scattering matrix $\mathbf{\Phi}\in\mathbb{C}^{M\times M}$. The scattering matrix generally describes the scattering characteristics of the $M$-port reconfigurable impedance network regardless of specific circuit designs, which relates the voltage of incident waves and reflected waves from the $M$ ports. As per the microwave network theory, for lossless reconfigurable impedance network, the scattering
matrix should be unitary, 
% $\mathbf{\Phi}^{H}\mathbf{\Phi}\preceq\mathbf{I}_{M}$, which denotes $\mathbf{I}_{M}-\mathbf{\Phi}^{H}\mathbf{\Phi}$ is positive semi-definite. Particularly, when the reconfigurable impedance network is lossless, we have a unitary constraint for the scattering matrix, 
that is the power of reflected waves is equal to that of the incident waves. It should be noted that the characteristics of the scattering matrix is associated with the circuit topology of the $M$-port reconfigurable impedance network. In this sense, in conventional RIS 1.0, each port
is connected to its own reconfigurable impedance without any connection across ports, referred to as single-connected RIS in \cite{shen2021}, which yields a diagonal scattering matrix. However, in BD-RIS, part of/all the ports are connected to each other so that the scattering matrix is not limited to being diagonal. In the following subsection, we will classify BD-RIS by the characteristics of scattering matrix, supported modes, and architectures.

\vspace{-0.4 cm}

\subsection{BD-RIS Classification}

\begin{figure*}
\centering{}\includegraphics[width=0.95\textwidth]{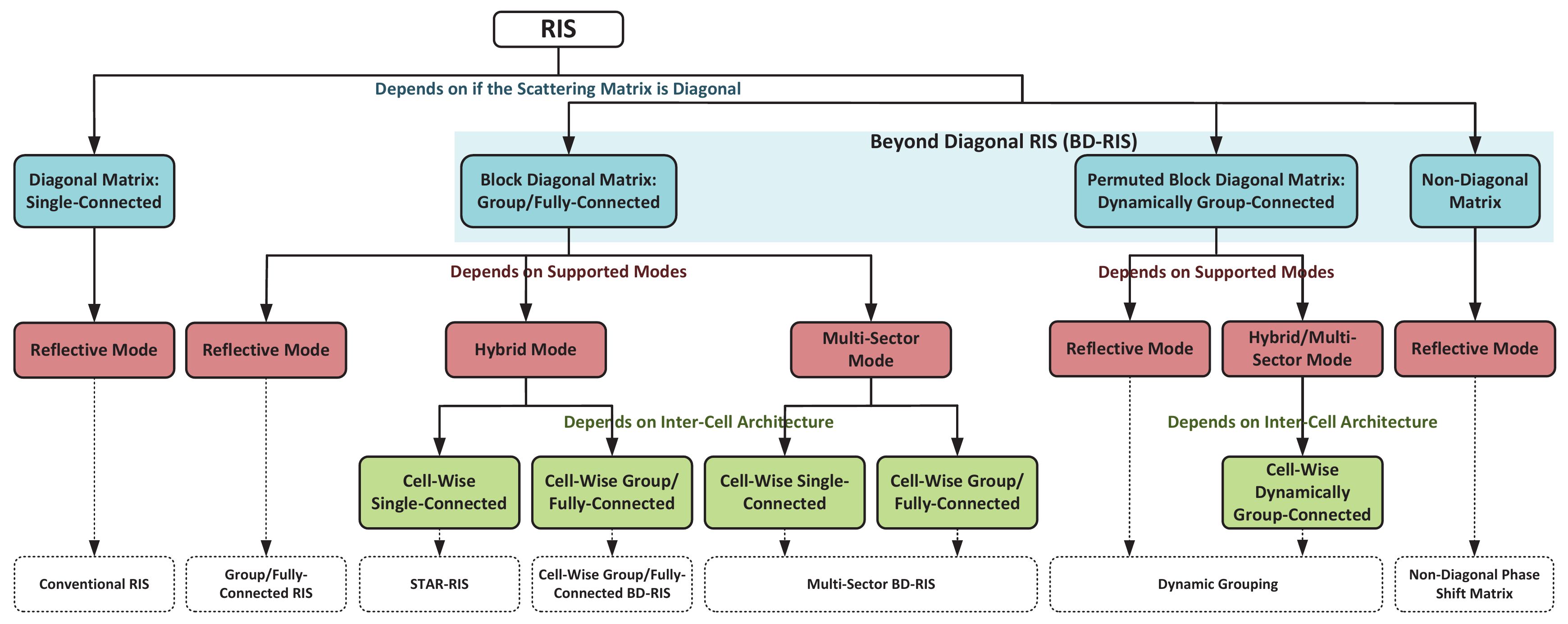}
\caption{RIS classification tree.}\vspace{-0.4 cm}
\label{fig:RIS_tree} 
\end{figure*}

We establish a three-layer RIS classification tree as shown in Fig. \ref{fig:RIS_tree}, where each layer is explained in detail as below.

The \textit{first} layer is classified by the characteristics of the scattering matrix $\mathbf{\Phi}$. \textit{1) Block Diagonal Matrix:} In this category, the $M$ antennas are uniformly divided into $G$
groups and antennas within the same group are connected to each other while those across groups are not connected. We refer to this category as group-connected RIS \cite{shen2021} and the corresponding scattering matrix $\mathbf{\Phi}$ is a block diagonal matrix with each block being unitary, which enables manipulating not only the phase but also the magnitude of incident waves and thus a better performance than the conventional RIS 1.0. Particularly, when there is only one group
$G=1$, i.e. all the $M$ antennas are connected to each other, it is referred to as fully-connected RIS \cite{shen2021}, which results in a unitary scattering matrix. Besides, the conventional RIS 1.0,
i.e. single-connected RIS, can be regarded as a special case of group-connected RIS with $M$ groups, which has a diagonal scattering matrix. 
\textit{2) Permuted Block Diagonal Matrix:} In this category, the grouping strategy, that is how the $M$ antennas are grouped, for the group-connected RIS is adaptive to the channel state information (CSI), which is thus referred to as dynamically group-connected RIS. The resulting scattering
matrix is a permuted block diagonal matrix \cite{li2022dynamic}, which provides higher flexibility in beam control than the fixed group-connected RIS. \textit{3) Non-Diagonal Matrix:} In this category, antennas are linked in pairs through phase shifters so that the signal impinging
on one antenna is purely reflected from another antenna, which results in an asymmetric non-diagonal  scattering matrix \cite{li2022reconfigurable} and a higher power gain than conventional RIS 1.0.

\begin{figure*}
    \centering{}\includegraphics[width=0.94\textwidth]{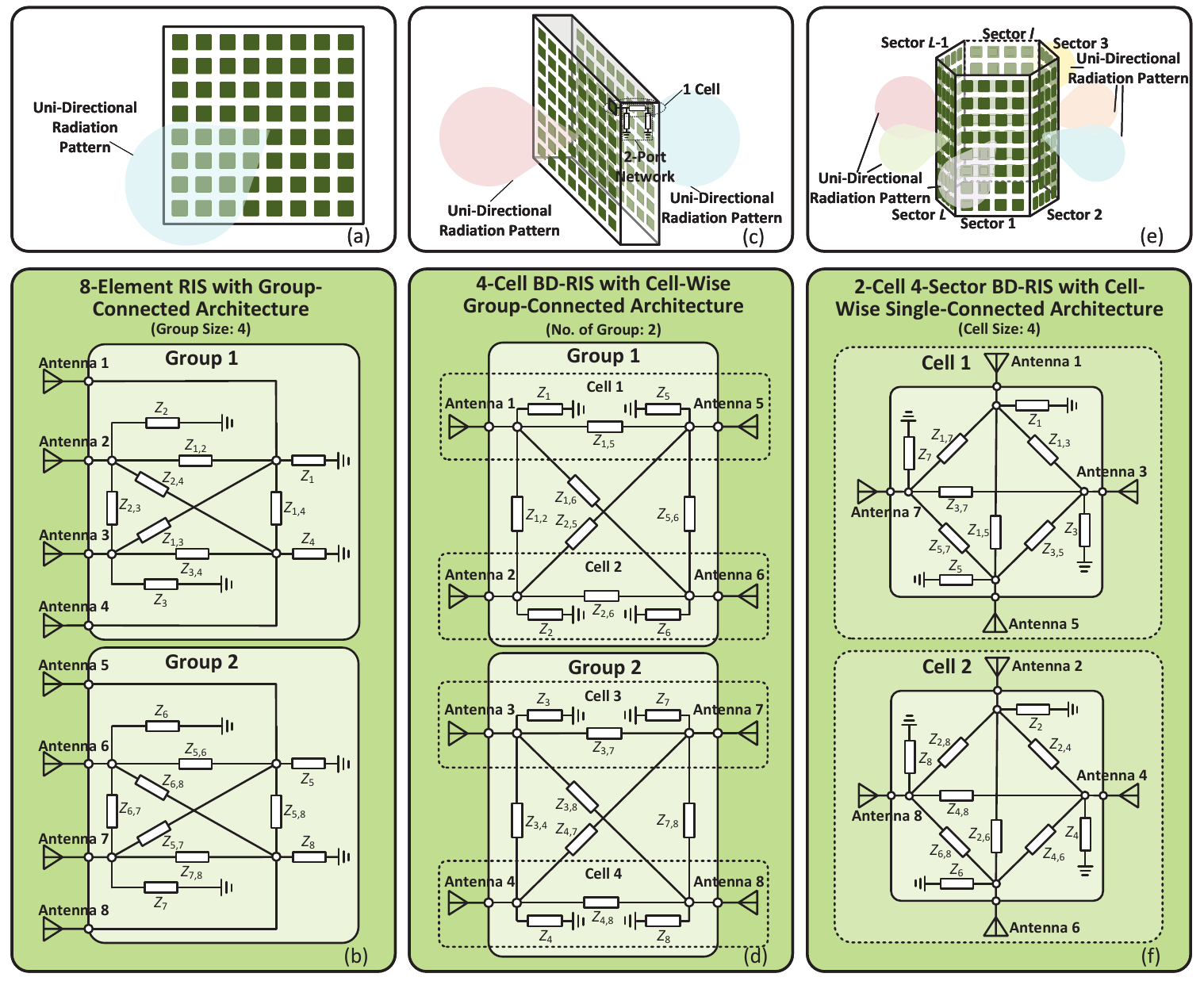}
    \caption{RISs with the same circuit topologies of reconfigurable impedance
    network while supporting different modes. (a) RIS with reflective
    mode and (b) group-connected architecture; (c) RIS with hybrid mode
    and (d) cell-wise group-connected architecture; (e) RIS with multi-sector
    mode and (f) cell-wise single-connected architecture.}\vspace{-0.4 cm}
    \label{fig:RIS_Architecture} 
\end{figure*}

The \textit{second} layer is classified by the modes supported by RIS, including reflective, hybrid, and multi-sector modes as detailed in the following. 
\textit{1) Reflective Mode:} In this mode, signals impinging on one side of the RIS are reflected toward the same side, yielding a half-space coverage. To support the reflective mode, all
the $M$ antennas of RIS are placed towards the same direction as shown in Fig. \ref{fig:RIS_Architecture}(a). Mathematically, the RIS with reflective mode is characterized by the matrix $\mathbf{\Phi}$ with a unitary constraint. 
\textit{2) Hybrid Mode:} In this mode, signals impinging on one side of the RIS can be partially reflected toward the same side and partially transmitted toward the opposite side, yielding a whole space coverage. The RIS with hybrid mode is also known as simultaneous transmitting and reflecting RIS (STAR-RIS) or intelligent omni-surface (IOS) \cite{zhang2022intelligent}.
To support the hybrid mode, each two antennas with uni-directional radiation pattern are back to back placed to form one \textit{cell}, and are connected to a 2-port fully-connected reconfigurable impedance network \cite{li2022} as shown in Fig. \ref{fig:RIS_Architecture}(c), so that each antenna in one cell respectively covers half space to achieve full-space coverage. Mathematically, the RIS with hybrid mode is characterized by two matrices, $\mathbf{\Phi}_{\mathrm{r}}\in\mathbb{C}^{\frac{M}{2}\times\frac{M}{2}}$ and $\mathbf{\Phi}_{\mathrm{t}}\in\mathbb{C}^{\frac{M}{2}\times\frac{M}{2}}$,
which satisfy that $\mathbf{\Phi}_{\mathrm{r}}^{H}\mathbf{\Phi}_{\mathrm{r}}+\mathbf{\Phi}_{\mathrm{t}}^{H}\mathbf{\Phi}_{\mathrm{t}}=\mathbf{I}_{\frac{M}{2}}$.
\textit{3) Multi-Sector Mode:} This mode is a generalization of hybrid mode. In this mode, the full space is divided into $L$ sectors ($L\geq2$) and signals impinging on one sector of RIS can be partially reflected toward the same sector and partially scattered toward the other $L-1$
sectors. 
To support the multi-sector mode, in each cell there are $L$ antennas placed at each edge of an $L$-sided polygon, with each antenna having a uni-directional radiation pattern covering $1/L$ to avoid overlapping among sectors, and the $L$ antennas are connected to an $L$-port fully-connected reconfigurable impedance network, as shown in Fig. \ref{fig:RIS_Architecture}(e).
Hence, the multi-sector mode can cover the full space as the hybrid mode, while providing higher performance gains than the hybrid mode, thanks to the use of higher-gain antennas with narrower beamwidth covering $1/L$ space.
Mathematically, the RIS with multi-sector mode is characterized by $L$ matrices, $\mathbf{\Phi}_{l}\in\mathbb{C}^{\frac{M}{L}\times\frac{M}{L}}$, $l=1,\ldots,L$, which satisfy $\sum_{l=1}^{L}\mathbf{\Phi}_{l}^{H}\mathbf{\Phi}_{l}=\mathbf{I}_{\frac{M}{L}}$.

The \textit{third} layer is classified by the inter-cell architecture, i.e., how the cells are connected to each other, in BD-RIS with hybrid/multi-sector modes. Analogous to the first layer in RIS classification tree, here we have cell-wise single/group/fully-connected architectures, where
the resulting $\mathbf{\Phi}_{\mathrm{r}}$ and $\mathbf{\Phi}_{\mathrm{t}}$ for hybrid mode or $\mathbf{\Phi}_{l}$ $\forall l$ for multi-sector mode are diagonal/block diagonal/full matrices, respectively. In \cite{li2022}, it is shown that the cell-wise group/fully connected architecture has a better performance than the cell-wise single connected architecture, i.e. the STAR-RIS/IOS. To further enhance the performance, we have cell-wise dynamically group-connected architecture, where the inter-cell
grouping strategy is adaptive to CSI and the resulting $\mathbf{\Phi}_{\mathrm{r}}$ and $\mathbf{\Phi}_{\mathrm{t}}$ for hybrid mode or $\mathbf{\Phi}_{l}$ $\forall l$ for multi-sector mode are permuted block diagonal matrices \cite{li2022dynamic}.

\vspace{-0.5 cm}

\subsection{Unified Architectures and Modes}

It should be noted that BD-RIS with different modes and architectures are realized by group-connected reconfigurable impedance network together with different antenna array arrangements. To get insights into the essence of BD-RIS with different modes/architectures, three examples are illustrated in Fig. \ref{fig:RIS_Architecture}, including 1) a BD-RIS with reflective mode and group-connected architecture, 2) a BD-RIS with hybrid mode and cell-wise group-connected architecture,
and 3) a BD-RIS with multi-sector mode and cell-wise single-connected architecture. From Figs. \ref{fig:RIS_Architecture}(b), (d), and (f), we can find these three BD-RISs have the same circuit topology
of reconfigurable impedance network but different antenna array arrangements, which results in different modes and inter-cell architectures. For clarity, we summarize the circuit complexity, that is the required number of reconfigurable impedance components, of BD-RIS with nine different modes/architectures in Table \ref{tab:complexity}.

\begin{table}[t]     
    \caption{Circuit Complexity of BD-RIS with Nine Modes/Architectures}     
    \centering      
    \begin{threeparttable}         
        \begin{tabular}{|c|c|c|c|}         
            \hline          
            \multirow{3}{*}{\diagbox[height = 27pt, innerleftsep = 4pt, innerrightsep = 1pt]{Mode}{Architecture\\(Inter-Cell)}} & Cell-Wise & Cell-Wise & Cell-Wise \\         
            \multirow{3}{*}{} & Single- & Group- & Fully- \\         
            \multirow{3}{*}{} & Connected\tnote{$\dagger$} & Connected\tnote{$\dagger$} & Connected\tnote{$\dagger$} \\         
            \hline         
            \hline          
            Reflective & $M$ & \multirow{3}{*}{$(\frac{M}{G}+1)\frac{M}{2}$} & \multirow{3}{*}{$(M+1)\frac{M}{2}$}\\         
            \cline{1-2}         
            Hybrid & $\frac{3}{2}M$ & \multirow{3}{*}{} & \multirow{3}{*}{}\\               
            \cline{1-2}         
            Multi-Sector & $(L+1)\frac{M}{2}$ & \multirow{3}{*}{} & \multirow{3}{*}{}\\          \hline          
        \end{tabular}  
        \begin{tablenotes}
            \footnotesize
            \item[$\dagger$] $M$: number of RIS elements; $G$: number of groups for reconfigurable impedance network; $L$: number of sectors for multi-sector BD-RIS.
        \end{tablenotes}      
    \end{threeparttable}\vspace{-0.4 cm}
    \label{tab:complexity}
\end{table} 

\vspace{-0.2 cm}

\section{Performance Evaluation for BD-RIS}

\begin{figure}
\includegraphics[scale=0.47]{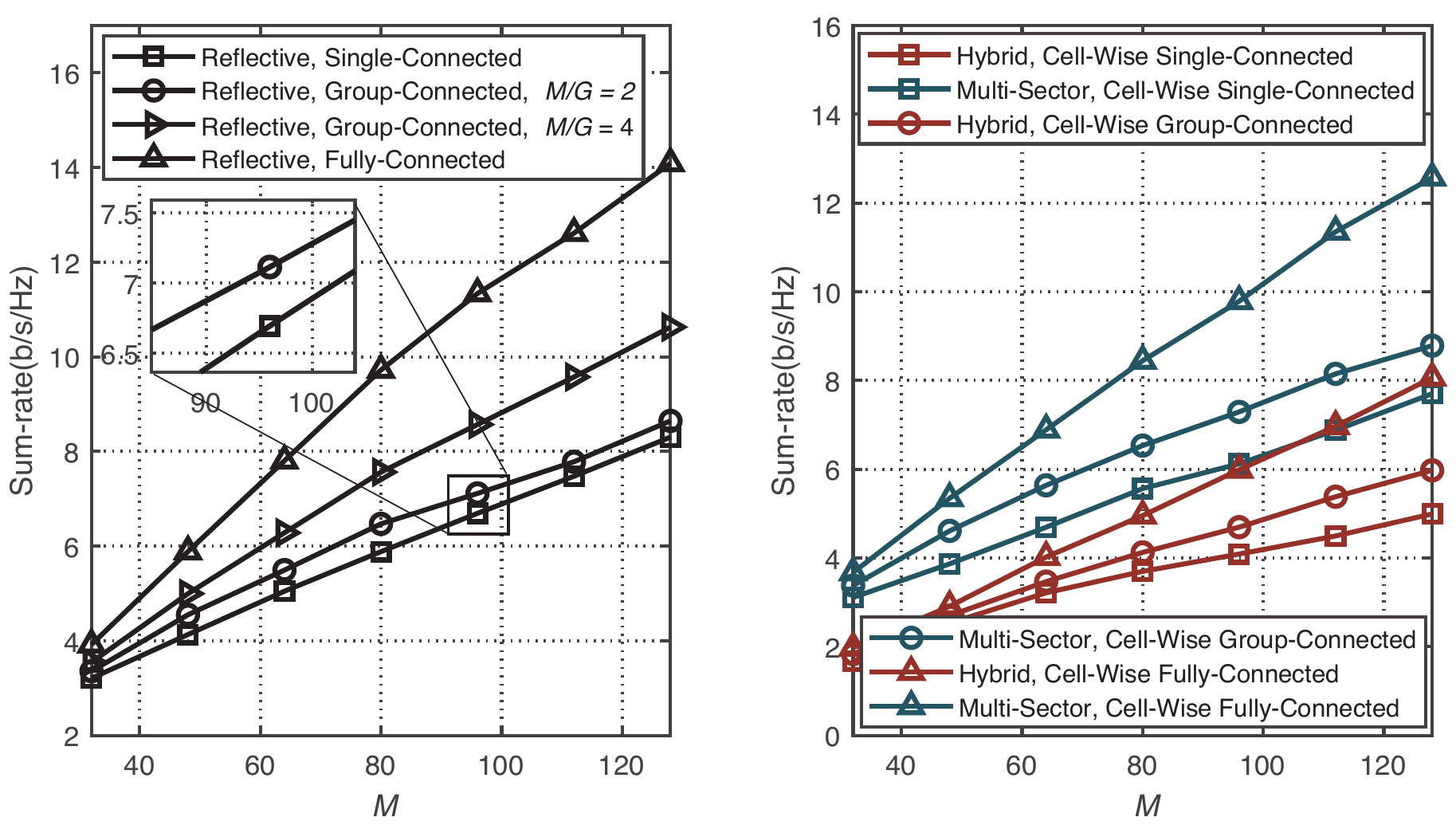}
\centering{}\caption{Sum-rate versus the number of BD-RIS antennas. Left: BD-RIS with
reflective mode; right: BD-RIS with hybrid/multi-sector modes.}\vspace{-0.4 cm}
\label{fig:SR_M}
\end{figure}

In this section, we evaluate the performance of BD-RIS with different modes and architectures. 
To that end, we consider a BD-RIS aided multiuser multiple input single output (MU-MISO) system, where a four-antenna transmitter serves four single-antenna users with the aid of BD-RIS. The four users are distributed evenly across the sectors covered by the BD-RIS, that is, four users on one side for the reflective mode, two users on either side for the hybrid mode, and one user within each sector for the multi-sector mode.
The transmit precoder and BD-RIS are jointly optimized to maximize the sum-rate of the MU-MISO system as detailed in \cite{li2022,li2022beyond}.
Fig. \ref{fig:SR_M} shows the sum-rate performance versus the number of BD-RIS antennas for the BD-RIS with nine different modes and architectures.
In Fig. \ref{fig:SR_M}, we consider a typical sub-6GHz narrowband scenario with carrier frequency 2.4 GHz.
In this scenario, the direct link between the transmitter and users is assumed to be blocked. 
The distance between the transmitter and the BD-RIS is set as 100 m. The distance between the BD-RIS and users is set as 10 m \cite{wu2021intelligent}. Results for other settings can be found in \cite{li2022,nerini2021reconfigurable}.
Channels from the transmitter to the BD-RIS and from BD-RIS to users are modeled as a combination of small-scale fading and large-scale fading. 
Specifically, the small-scale fading components follow the Rician fading with Rician factor 0 dB.
As per the results in \cite{shen2021,li2022}, decreasing the Rician factor yields higher-rank channels, which provide more degree of freedom for the design of BD-RIS with group/fully-connected architectures and thus can further enhance the performance compared with conventional RIS 1.0.
The large-scale fading components are related to the BD-RIS antenna gains and path loss, which are modeled and calculated based on \cite{li2022beyond}.
Transmit power is set as $P=30$ dBm. The noise power at each user is set as $-80$ dBm. The group size $M/G$ for hybrid and multi-sector modes are respectively set as 4 and 8. 
For fair comparison, we fix the number of BD-RIS antennas and coverage of the hybrid and multi-sector BD-RIS, such that the number of antennas for each sector decreases while the antenna gain increases with $L$ to provide higher channel gain.
We make the following observations.

\textit{First}, under the reflective mode, BD-RIS with group/fully-connected architectures always achieves better performance than conventional RIS 1.0 due to the more general constraint of the BD-RIS matrix. 
According to \cite{li2022}, the optimization complexity for BD-RIS design is $\mathcal{O}(IM^3/G^2)$, where $I$ denotes the number of iterations.
Therefore, combining the circuit complexity in Table \ref{tab:complexity}, the optimization complexity for BD-RIS design, and the results in Fig. \ref{fig:SR_M}, we observe that the BD-RIS with group size 2 achieves a good trade-off between performance and complexity by increasing the sum-rate by around 8\% at the expense of half more impedance components and around 4 times the optimization complexity.

\textit{Second}, with the same cell-wise architecture, the BD-RIS with multi-sector mode always outperforms that with hybrid mode, even though the number of antennas covering each user for the former case is reduced compared to the latter. 
This is because the BD-RIS antennas with multi-sector mode has narrower beamwidth compared to those with
hybrid mode, and thus provide higher gains. 
More interestingly, multi-sector BD-RIS with cell-wise single-connected architecture outperforms the
hybrid BD-RIS with inter-cell single/group-connected architectures. 
This finding implies that with proper antenna array arrangements of BD-RIS, a reduced circuit complexity can achieve both satisfactory performance and full-space coverage.

\textit{Third}, for all three modes, the sum-rate achieved by BD-RIS with (cell-wise) fully-connected architectures grows faster with $M$ than that with single-connected architecture. This phenomenon can be explained as follows: BD-RIS with fully-connected architectures mathematically results in BD-RIS matrices with a larger number of non-zero elements, which provides higher flexibility of passive beamforming. 
It should also be noted that the increased design flexibility of BD-RIS with fully-connected architectures is achieved at the expense of increasing circuit complexity as summarized in Table \ref{tab:complexity}, which indicates that the circuit complexity of BD-RIS grows linearly with $M$ for single-connected architecture, but grows quadratically with $M$ for fully-connected architectures.

\section{Benefits and Potential Applications of BD-RIS}

\begin{figure*}
    \centering{}
    \includegraphics[width=0.9\textwidth]{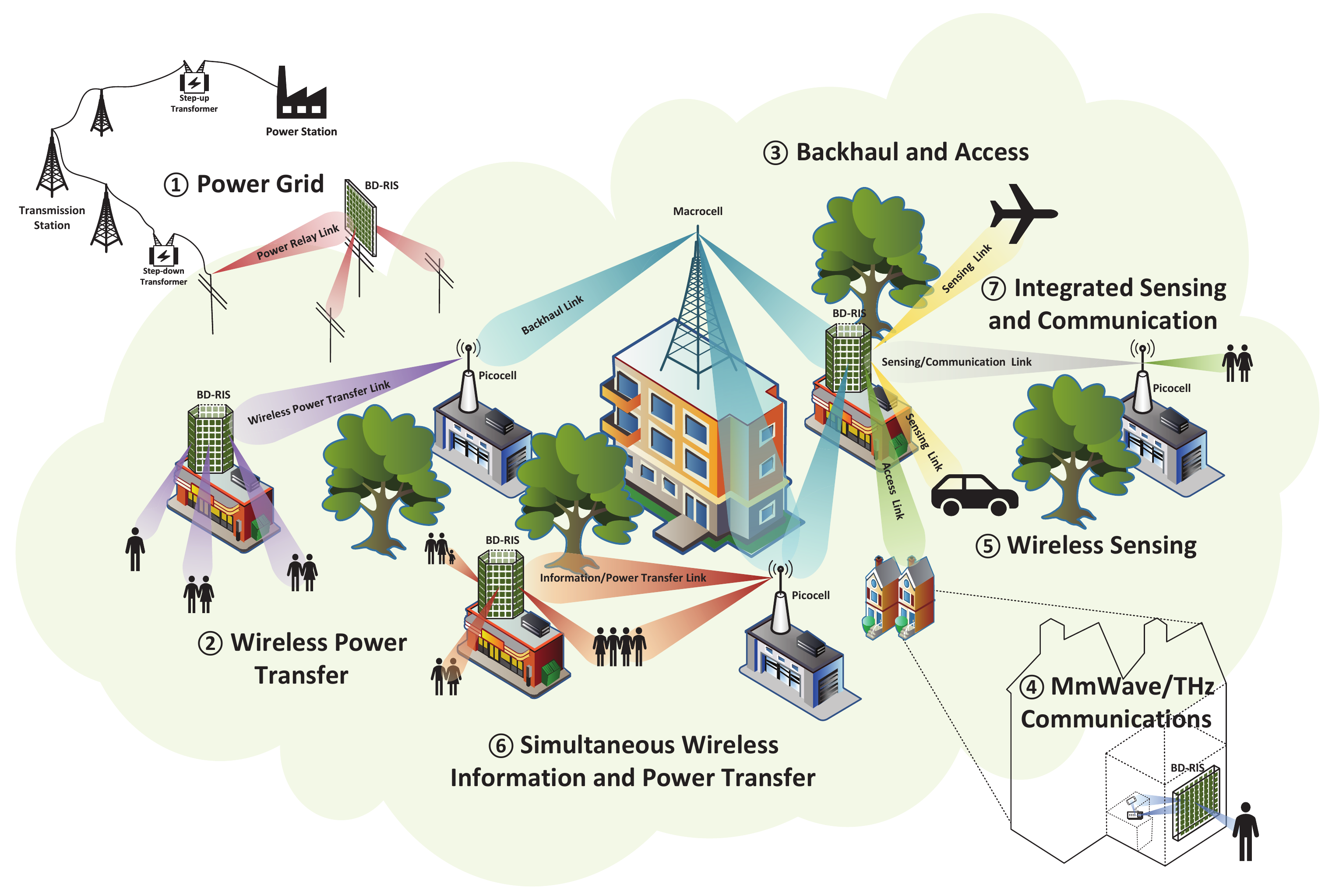}
    \caption{Potential applications of BD-RIS. \ding{172} BD-RIS works as a passive relay in the power grid; \ding{173} BD-RIS insists wireless power transfer; \ding{174} BD-RIS enables wireless backhaul and access; \ding{175} BD-RIS insists  millimeter wave (mmWave)/Terahertz (THz)
    communications; \ding{176} BD-RIS insists wireless sensing; \ding{177} BD-RIS insists simultaneous wireless information and power transfer; \ding{178} BD-RIS enables integrated sensing and communication.}\vspace{-0.3 cm}
    \label{fig:BD_RIS_App} 
\end{figure*}

We have shown the pronounced benefits of BD-RIS compared to conventional RIS 1.0 in the example of MU-MISO system in Section III. In this section, we summarize the key benefits of BD-RIS and discuss potential applications of BD-RIS in various wireless systems as illustrated in Fig. \ref{fig:BD_RIS_App}.

\subsection{Benefits of BD-RIS}

\subsubsection{High Flexibility in Wave Manipulation}

Compared with the conventional RIS 1.0 which can only manipulate the diagonal entries of the scattering matrix, the BD-RIS has higher flexibility in manipulating both diagonal and off-diagonal entries of the scattering matrix, which further boosts the performance in various wireless systems. Results in \cite{shen2021} show that BD-RIS with reflective mode and group/fully-connected architectures increases the received power by up to 62\% compared to conventional RIS 1.0; results in \cite{santamaria2023snr} show that BD-RIS with reflective mode and fully-connected architectures achieves 2 dB of signal-to-noise ratio (SNR) gain compared to conventional RIS 1.0; results in \cite{wang2023channel} show that BD-RIS relying on non-reciprocal circuits can be maliciously used to break the uplink-downlink channel reciprocity; results in \cite{bartoli2023spatial} show that BD-RIS achieves better rate performance than conventional RIS 1.0 in near-field line of sight (LoS) scenarios.
% results in \cite{li2022} show that hybrid BD-RIS with group/fully-connected architectures achieves up to 75\% higher sum-rate than STAR-RIS.

\subsubsection{Full-Space Coverage}

Compared with conventional RIS 1.0 which can only cover half-space, the BD-RIS utilizing appropriate group-connected reconfigurable impedance network and antenna array arrangement can support the hybrid and multi-sector modes to realize full-space coverage \cite{li2022}, \cite{li2022beyond}.
Moreover, the multi-sector mode can provide high channel gain thanks to the narrower beamwidth and higher gain of each RIS antenna, and thus effectively extend the communication range for full-space coverage \cite{li2022beyond}.

\subsubsection{Facilitating Deployments}

BD-RIS with hybrid and multi-sector modes facilitates practical deployments.
Benefiting from the full-space coverage, the locations of the BD-RIS could be more flexible than conventional RIS 1.0.

\subsubsection{Low Complexity in Resolution Bit Number}

When considering RIS with discrete values, BD-RIS is shown to achieve a better performance than conventional RIS 1.0 with fewer resolution bits \cite{nerini2021reconfigurable}, due to the high flexibility of reconfigurable impedance network. 
Specifically, results in \cite{nerini2021reconfigurable} show that to achieve satisfactory performance close to the continuous-value case, four resolution bits are required in conventional RIS 1.0, but only one resolution bit is sufficient in fully-connected BD-RIS with reflective mode.
Such reduction of resolution bits is beneficial for implementation of BD-RIS.

\subsubsection{Low Complexity in Element Number}

As the BD-RIS, especially with multi-sector mode, greatly enhances the performance in various wireless networks, given the same performance requirement, the required BD-RIS element number can be effectively
reduced. 
Results in \cite{li2022beyond} show that a 6-sector BD-RIS can maintain the same sum-rate as a 3-sector BD-RIS with a number of elements reduced by 20\%.
This benefit lowers the RIS complexity, cost, and form factor.

\subsection{Potential Applications of BD-RIS}

\subsubsection{Wireless Power Relay/Transfer}

\begin{figure}
    \centering{}
    \includegraphics[width=0.47\textwidth]{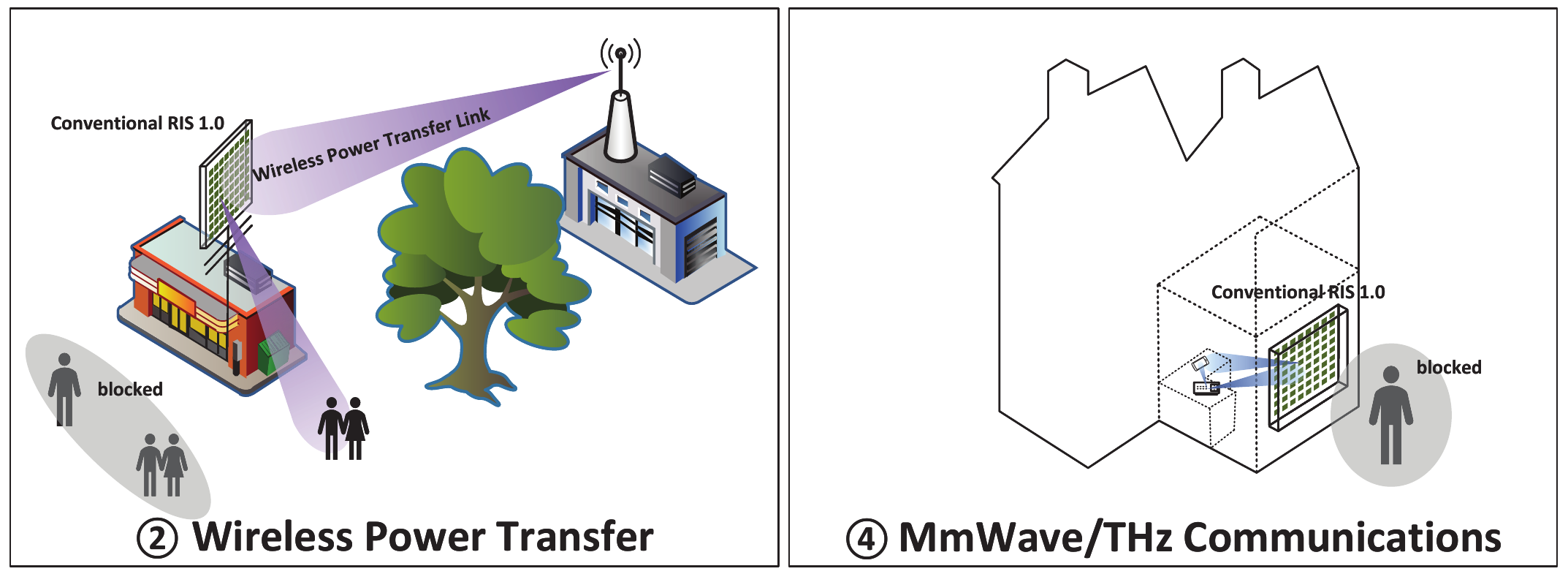}
    \caption{Applications of conventional RIS 1.0. \ding{173} Conventional RIS 1.0 insists wireless power transfer, where part of receivers is out of converage; \ding{175} Conventional RIS 1.0 insists  millimeter wave (mmWave)/Terahertz (THz) communications, where the receiver outside the house is blocked.}\vspace{-0.3 cm}
    \label{fig:RIS10_App} 
\end{figure}

One promising application of BD-RIS is to deploy it in the power grid to relay wireless power. 
% The power grid is an electricity system which is generally used to carry power from a few central generators to numerous users/customers/devices. Specifically, the power grid consists of the power generation, the transmission grid which moves the up-stepped power over long distances to substations, and the distribution grid which delivers the down-stepped power to serve users \cite{fang2011smart}.
In Fig. \ref{fig:BD_RIS_App} we provide a diagram of employing BD-RIS in the power distribution grid to relay wireless power. 
For comparison, we illustrate the case of employing conventional RIS 1.0 in the same scenario. Combining Figs. \ref{fig:BD_RIS_App} and \ref{fig:RIS10_App}, we observe that with proper power levels, suitable deployments and locations of BD-RIS, the BD-RIS
could effectively aid the wireless power transfer, guaranteeing that the power can be transferred to all receivers. However, applying conventional RIS 1.0 can only guarantee the service for specific directions.

\subsubsection{Wireless Communications}

Another interesting application of BD-RIS is to enable flexible and scalable integrated access and backhaul (IAB) \cite{madapatha2020integrated}.
% IAB is one of the promising techniques for 5G networks, where the operator can use part of the radio resources for wireless backhauling while providing the existing cellular services in the same node. 
Fig. \ref{fig:BD_RIS_App} illustrates the BD-RIS assisted IAB, where the BD-RIS can be flexibly deployed in the IAB system to not only assist the wireless backhauling between the macrocell and picocells, but also the wireless access between picocells and users. 
Specifically, the wireless backhauling usually have complicated propagation environments
and various obstacles, e.g. trees and high buildings as shown in Fig. \ref{fig:BD_RIS_App}.
BD-RIS with full space coverage and high gain performance can be easily incorporated into real  environments to bypass the obstacles and assist/enhance the wireless backhaul. 
Meanwhile, wireless access, especially in millimeter wave or Terahertz wireless frequencies, usually has sparse and highly-directional channels, suffers from high path loss, and is vulnerable to blockages. 
In this case, BD-RIS is more appealing in providing highly-directional beams to align with low-rank channels, compensate for the severe path loss, and enlarge coverage.
Specifically, when the transmitter is inside the house while receivers are both inside and outside the house as illustrated in Figs. \ref{fig:BD_RIS_App} and \ref{fig:RIS10_App}, BD-RIS enables joint indoor and outdoor communications thanks to the enhanced coverage, which cannot be achieved by conventional RIS 1.0.

\subsubsection{Wireless Sensing}

BD-RIS can also be deployed to boost the wireless sensing performance, such as improving the target detection accuracy and reducing the parameter estimation error, for targets enjoying line of sight (LoS) links.
More importantly, for those complicated propagation environments without LoS links between the radar and targets, such as the vehicle networks, BD-RIS enables wireless sensing and enlarges coverage by creating effective LoS links.

\subsubsection{Integrated Wireless Power Transfer, Communications, and Sensing}

In addition to stand-alone wireless power transfer, communications, and sensing, BD-RIS can also be used to assist integrated systems, such as simultaneous wireless information and power transfer
or integrated sensing and communication, as shown in Fig. \ref{fig:BD_RIS_App}. 
This can be effectively done with reasonable numbers of BD-RISs and simple deployments, while the network would be very complicated (or even not practical) using only conventional RIS 1.0.

Not limited to these applications, BD-RIS can be applied in all the conventional RIS 1.0 enabled systems, but with higher flexibility and better performance in architecture design, beam manipulation, and deployment than conventional RIS 1.0.

\section{Challenges and Future Work of BD-RIS}

While the BD-RIS has benefits compared with conventional RIS 1.0, there exist challenges in designing and implementing BD-RIS for practical wireless networks, which shed light on future research directions
for BD-RIS. In this section, we list five challenges and future work from the perspectives of hardware implementation, architecture design, RF impairments, channel estimation, and wideband modeling as follows.

\vspace{-0.2 cm}

\subsection{Hardware Implementation}

The hardware implementation of BD-RIS is a fundamental issue. 
Currently, only RIS with STAR-RIS/IOS \cite{zhang2022intelligent} has been implemented, which indicates the physical availability of modeling RIS as antenna arrays connected to reconfigurable impedance network, while the comprehensive hardware implementation of BD-RIS with different architectures is still on its way.
As per the model in Section II, an $M$-element BD-RIS consists of two parts and can be implemented as follows.

\subsubsection{$M$-Antenna Array}

For the reflective mode, we can use the conventional uniform linear or planar antenna array. For the hybrid mode, we need to place each two antennas with uni-directional radiation pattern (e.g. patch antenna) back to back to form a cell and then arrange all the cells in a uniform array. Furthermore, for the multi-sector mode, we need to place each $L$ antennas with narrow beamwidth at each edge of an $L$-side polygon to form a cell and arrange the cells in a uniform array.

\subsubsection{$M$-Port Reconfigurable Impedance Network}

As shown in Section II-C, the group-connected reconfigurable impedance network is the key to implement the BD-RIS with different modes and architecture. We can utilize tunable inductance and capacitance, e.g. varactors, to construct the group-connected reconfigurable impedance network as per the circuit topology shown in Section II-C, so that the continuous value BD-RIS can be implemented. Alternatively, we can use PIN diodes as switches to reconfigure the impedance network to implement discrete value BD-RIS. 
However, as the group size increases, the circuit complexity and cost also increase, which is unique in BD-RIS due to the inter-element connections that conventional RIS 1.0 does not have.
Hence, it is challenging but worthwhile to explore new BD-RIS architectures to achieve better performance-complexity trade-off.
The recent work \cite{nerini2023beyond} studies the conditions of BD-RIS connections which theoretically achieve the best performance-complexity trade-off using graph theory. Based on the derived conditions, two novel BD-RIS architectures, namely tree- and forest-connected, are proposed to reach the trade-off.  
Nevertheless, the practical limitation of those novel architectures remains unexplored, thereby inducing new research directions: implementing and prototyping different BD-RIS architectures taking into account practical hardware issues to verify their superiority compared to conventional RIS 1.0.

\vspace{-0.2 cm}

\subsection{RF Impairments}
Most existing designs for RIS 2.0 focus on idealized models with perfect matching and no mutual coupling of antennas, and lossless impedance components with continuous values. 
However, those idealized assumptions do not always hold in practical scenarios, thereby generating the following twofold challenges. 1) When mismatching and mutual coupling exist in practical scenarios, the channel model will no longer be a linear function of $\mathbf{\Theta}$. This issue will complicate the beamforming design, which has never been investigated in the existing works. 2) When using PIN diodes to implement the discrete value BD-RIS, it is not possible to design discrete values of the BD-RIS matrix using simple quantizations as in conventional RIS 1.0 since entries of the BD-RIS matrix depend on each other.
Therefore, it is challenging but important to take into account the RF impairments in the BD-RIS design.
%  For conventional RIS 1.0 with diagonal phase shift matrix, the discretization is straightforward by uniformly sampling the phase within $2\pi$. However, for BD-RIS
% with unitary matrix, the discretization is difficult. 
% Instead, we need to determine the discrete values of the reactance matrix for
% the reconfigurable impedance network. 
In \cite{nerini2021reconfigurable}, a potential direction for the codebook design of group/fully-connected BD-RIS with reflective mode has been provided. 
Nevertheless, investigating 
discrete value BD-RIS design with hybrid/multi-sector modes and different architectures still remains an open problem.
% Therefore, one of the meaningful reseach directions is to develop efficient beamforming design approaches while accurately capturing the RF impairments of BD-RIS.

\vspace{-0.2 cm}

\subsection{Channel Estimation}

The pronounced performance gain brought by the BD-RIS requires accurate CSI. For conventional RIS 1.0, there are two channel estimation strategies:
1) Semi-passive channel estimation by equipping a few low-power RF chains to the RIS to enable the pilot transmission/reception; 
2) Pure passive channel estimation by estimating the cascaded transmitter-RIS-user
channels with pre-defined RIS patterns, which characterize the variation of RIS matrix during the training period. 
The first channel estimation strategy is still available for the proposed BD-RIS but at the expense of additional power consumption due to the introduced RF chains.
% However, these channel estimation strategies cannot be directly applied in BD-RIS. 
% For the first strategy, we need to reconsider the deployment of RF chains and the pilot design in the channel estimation process due to the different architectures/modes of BD-RIS. 
The second strategy utilized in conventional RIS 1.0 cannot be directly used for BD-RIS since the dimension of the cascaded channel depends on the circuit topology of the reconfigurable impedance network.
Thus, it is important to develop new channel estimation strategies with reduced power consumption for BD-RIS in the near future.

\subsection{Wideband BD-RIS Modeling}

The current BD-RIS model is only for narrowband communication. When it comes to wideband communications, the modeling of BD-RIS should take into account the frequency response of the reconfigurable impedance network. 
Specifically, each reconfigurable component of the impedance network is frequency dependent, where the frequency response is determined by the circuit designs. 
Moreover, different reconfigurable components are related to each other due to the inter-element connection in BD-RIS, which makes it impossible to do element-by-element design as in conventional RIS 1.0. This motivates the exploration of simpler wideband modeling and effective beamforming design for BD-RIS.
% Consequently, the resulting BD-RIS matrices at different frequencies are dependent on each other, which will complicate the wideband BD-RIS design. 
% To tackle the frequency dependent BD-RIS matrices and simplify the wideband BD-RIS design, a possible solution is to 1) analyze and fit the relationship between amplitudes/phase shifts of BD-RIS matrices and frequencies based on practical and specific circuits and 2) consider the wideband BD-RIS design based on the fitted frequency dependent BD-RIS model.

\section{Conclusion}

\label{sc:Conclusion}

In this paper, we depart from conventional RIS 1.0 with diagonal phase shift matrices and branch out to  RIS 2.0 (BD-RIS) with beyond diagonal scattering matrices. 
Specifically, we model and classify the BD-RIS based on fundamental circuit topologies of reconfigurable impedance network. 
In addition, we highlight the benefits of BD-RIS with different modes/architectures in providing high flexibility in wave manipulation, achieving full-space coverage, flexibility in various deployments,
and low complexity in resolution bit and element numbers of the impedance network. 
Potential applications, challenges, and future work of BD-RIS are also discussed and summarized. 
As BD-RIS is a brand-new advance in RIS technology that remains unexplored from various perspectives,
it is hoped that this paper could offer a useful and stimulating guide on future research directions of BD-RIS.

\bibliographystyle{IEEEtran}
\bibliography{references}

\begin{thebibliography}{10}
\providecommand{\url}[1]{#1}
\csname url@samestyle\endcsname
\providecommand{\newblock}{\relax}
\providecommand{\bibinfo}[2]{#2}
\providecommand{\BIBentrySTDinterwordspacing}{\spaceskip=0pt\relax}
\providecommand{\BIBentryALTinterwordstretchfactor}{4}
\providecommand{\BIBentryALTinterwordspacing}{\spaceskip=\fontdimen2\font plus
\BIBentryALTinterwordstretchfactor\fontdimen3\font minus
  \fontdimen4\font\relax}
\providecommand{\BIBforeignlanguage}[2]{{%
\expandafter\ifx\csname l@#1\endcsname\relax
\typeout{** WARNING: IEEEtran.bst: No hyphenation pattern has been}%
\typeout{** loaded for the language `#1'. Using the pattern for}%
\typeout{** the default language instead.}%
\else
\language=\csname l@#1\endcsname
\fi
#2}}
\providecommand{\BIBdecl}{\relax}
\BIBdecl

\bibitem{di2020smart}
M.~Di~Renzo, A.~Zappone, M.~Debbah, M.-S. Alouini, C.~Yuen, J.~De~Rosny, and
  S.~Tretyakov, ``Smart radio environments empowered by reconfigurable
  intelligent surfaces: How it works, state of research, and the road ahead,''
  \emph{IEEE J. Sel. Areas Commun.}, vol.~38, no.~11, pp. 2450--2525, 2020.

\bibitem{wu2019towards}
Q.~Wu and R.~Zhang, ``Towards smart and reconfigurable environment: Intelligent
  reflecting surface aided wireless network,'' \emph{IEEE Commun. Mag.},
  vol.~58, no.~1, pp. 106--112, 2019.

\bibitem{wu2021intelligent}
Q.~Wu, S.~Zhang, B.~Zheng, C.~You, and R.~Zhang, ``Intelligent reflecting
  surface-aided wireless communications: A tutorial,'' \emph{IEEE Trans.
  Commun.}, vol.~69, no.~5, pp. 3313--3351, 2021.

\bibitem{shen2021}
S.~Shen, B.~Clerckx, and R.~Murch, ``Modeling and architecture design of
  reconfigurable intelligent surfaces using scattering parameter network
  analysis,'' \emph{IEEE Trans. Wireless Commun.}, vol.~21, no.~2, pp.
  1229--1243, 2022.

\bibitem{li2022dynamic}
H.~Li, S.~Shen, and B.~Clerckx, ``A dynamic grouping strategy for beyond
  diagonal reconfigurable intelligent surfaces with hybrid transmitting and
  reflecting mode,'' \emph{IEEE Trans. Veh. Technol.}, 2023.

\bibitem{li2022reconfigurable}
Q.~Li, M.~El-Hajjar, I.~A. Hemadeh, A.~Shojaeifard, A.~Mourad, B.~Clerckx, and
  L.~Hanzo, ``Reconfigurable intelligent surfaces relying on non-diagonal phase
  shift matrices,'' \emph{IEEE Trans. Veh. Technol.}, vol.~71, no.~6, pp.
  6367--6383, 2022.

\bibitem{li2022}
H.~Li, S.~Shen, and B.~Clerckx, ``Beyond diagonal reconfigurable intelligent
  surfaces: From transmitting and reflecting modes to single-, group-, and
  fully-connected architectures,'' \emph{IEEE Trans. Wireless Commun.},
  vol.~22, no.~4, pp. 2311--2324, 2023.

\bibitem{li2022beyond}
------, ``Beyond diagonal reconfigurable intelligent surfaces: A multi-sector
  mode enabling highly directional full-space wireless coverage,'' \emph{IEEE
  J. Sel. Areas Commun.}, vol.~41, no.~8, pp. 2446--2460, 2023.

\bibitem{santamaria2023snr}
I.~Santamaria, M.~Soleymani, E.~Jorswieck, and J.~Guti{\'e}rrez, ``{SNR}
  maximization in beyond diagonal {RIS}-assisted single and multiple antenna
  links,'' \emph{IEEE Signal Process. Lett.}, 2023.

\bibitem{wang2023channel}
H.~Wang, Z.~Han, and L.~Swindlehurst, ``Channel reciprocity attacks using
  intelligent surfaces with non-diagonal phase shifts,''
  \emph{arXiv:2309.11665}, 2023.

\bibitem{bartoli2023spatial}
G.~Bartoli, A.~Abrardo, N.~Decarli, D.~Dardari, and M.~Di~Renzo, ``Spatial
  multiplexing in near field {MIMO} channels with reconfigurable intelligent
  surfaces,'' \emph{IET Signal Process.}, vol.~17, no.~3, p. e12195, 2023.

\bibitem{nerini2021reconfigurable}
M.~Nerini, S.~Shen, and B.~Clerckx, ``Discrete-value group and fully connected
  architectures for beyond diagonal reconfigurable intelligent surfaces,''
  \emph{IEEE Trans. Veh. Technol.}, 2023.

\bibitem{zhang2022intelligent}
H.~Zhang and B.~Di, ``Intelligent omni-surfaces: Simultaneous refraction and
  reflection for full-dimensional wireless communications,'' \emph{IEEE Commun.
  Surveys \& Tutorials}, 2022.

\bibitem{madapatha2020integrated}
C.~Madapatha, B.~Makki, C.~Fang, O.~Teyeb, E.~Dahlman, M.-S. Alouini, and
  T.~Svensson, ``On integrated access and backhaul networks: Current status and
  potentials,'' \emph{IEEE Open J. Commun. Society}, vol.~1, pp. 1374--1389,
  2020.

\bibitem{nerini2023beyond}
M.~Nerini, S.~Shen, H.~Li, and B.~Clerckx, ``Beyond diagonal reconfigurable
  intelligent surfaces utilizing graph theory: Modeling, architecture design,
  and optimization,'' \emph{arXiv:2305.05013}, 2023.

\end{thebibliography}

\end{document}